# Inefficiencies in Digital Advertising Markets

Brett R Gordon, Northwestern University
Kinshuk Jerath, Columbia University
Zsolt Katona*, University of California, Berkeley
Sridhar Narayanan, Stanford University
Jiwoong Shin, Yale University
Kenneth C Wilbur*, University of California, San Diego

February 2020

Abstract:
Digital advertising markets are growing and attracting increased scrutiny. This paper explores four market inefficiencies that remain poorly understood: ad effect measurement, frictions between and within advertising channel members, ad blocking and ad fraud. These topics are not unique to digital advertising, but each manifests in new ways in markets for digital ads. We identify relevant findings in the academic literature, recent developments in practice, and promising topics for future research.

Acknowledgements:
We are grateful for many helpful suggestions from the special issue editors, John Deighton, Carl Mela and Chris Moorman, and four anonymous reviewers. We also received helpful comments from Ron Berman, Sundar Bharadwaj, Garrett Johnson, Mingyu Joo, Don Lehmann, Matt McGranaghan, Oded Netzer, Dave Reibstein, Catherine Tucker, Sri Venkataraman, Michel Wedel and Yi Zhu. Any remaining errors are ours alone.

Disclosure:
Gordon was previously a contractor and is currently an employee at Facebook; he donates all resulting income to charity. Authors have no other funding or conflicts of interest to report.

* Corresponding authors: zskatona@berkeley.edu, kcwilbur@ucsd.edu

Digital advertising markets have offered unprecedented innovations to marketers. Businesses can now advertise to finely targeted sets of individuals with customized commercial messages at specific locations and times in a variety of formats. Compared to traditional advertising, digital ads promise better targeting and relevance, personalized ad content, programmatic sales based on real-time auctions, and measurement of the co-occurrence of individual consumer ad exposures with a variety of online and offline response behaviors. These features have fundamentally altered marketers' spending: digital advertising revenues reached $108B in 2018, up 117% over 2014 (IAB 2019), with expectations to grow 19% and surpass cumulative traditional advertising revenues in 2019 (eMarketer 2019a).

Digital advertising markets sell a wide variety of search and display advertising opportunities to marketers. Although digital advertising is 25 years old, market structures are still changing rapidly. For example, a census of marketing technology firms showed an increase from 150 in 2011 to 7,040 in 2019 (Brinker 2019). The IAB Tech Lab recently introduced a series of broad-based initiatives, including a new real-time bidding standard, and Google moved from second-price to first-price auctions for display ads. Publishers have introduced a variety of ad formats, with spending typically following consumer attention and media usage: after initial growth in desktop display and search, recent growth has been more concentrated in social networking, video, audio, and mobile ads.

Yet there are indicators that unregulated markets for digital advertising have experienced problems. The E.U. has fined Google more than $9 billion in three antitrust cases and the U.S. Federal Trade Commission fined Facebook $5 billion after it broke a 2012 consent order (Case 19-cv-2184). Prominent politicians have criticized the industry and proposed structural reforms.

New privacy laws mandate transparency and consent requirements for data-driven advertising and user identification practices.

Several comprehensive reviews of digital platform markets have advised new regulation. For example, the Australian Competition and Consumer Commission (2019) made 23 recommendations, including that "a specialist digital platforms branch be established" to proactively monitor digital markets, enforce laws, conduct inquiries and recommend actions to address consumer harm and market failure. The U.K. House of Lords Select Committee on Communications (2019) reached similar conclusions, noting "a lack of understanding among policy-makers." Similar conclusions have been reached by the European Commission Directorate-General for Competition; the U.K. Department of Digital Culture, Media & Sport; and the U.K. Digital Competition Expert Panel; with ongoing investigations by the U.S. Federal Trade Commission and the U.K. Competition and Markets Authority, among others.

The purpose of this article is to review four prominent features that may limit digital advertising markets from reaching their maximum allocative efficiency (i.e., the extent to which markets distribute digital advertising opportunities to the agents that value the opportunities most, as in Harberger 1954).[1] The four topics are as follows.

- *Ad effect measurement* is the estimation of incremental effects of advertisements on consumer behaviors. Ignorance or uncertainty about ad effects may inefficiently distort advertisers' reservation prices, demand, budgets and bids for ads.

---

[1] We focus on allocative efficiency because it is a classic and fundamental measure of market performance, and is a frequent criterion economists use to recommend and evaluate government policies. We do not intend to discount other policy objectives such as consumer welfare, distributional concerns, competitiveness or innovation.

- *Organizational inefficiencies* occur within advertising organizations and between advertisers and their self-interested agencies, and may lead to inefficient advertising decisions.

- *Ad blocking* is a consumer technology that prevents ads from being displayed. Ad blockers may inefficiently appropriate advertising revenues and harm publishers' incentives to provide content.

- *Ad fraud* is a collection of practices that misrepresent advertising inventory or disguise machines as humans in order to steal advertising expenditures. Most industry estimates indicate that fraud takes 10-30% of total digital advertising revenue.

These features are important: 75% of brand marketers indicated ad effect measurement to be the leading "threat to digital ad budgets in 2019," with 69% of agency professionals reporting ad fraud as the leading threat (Benes 2019). We dedicate one section to each topic, selectively describing relevant scientific literature and credible industry knowledge, and identifying opportunities for future research. Although we discuss each issue in isolation, they may interact to exacerbate some of the inefficiencies we describe. We close with a broader discussion of policy-relevant research opportunities.

It is important to note that these four inefficiencies predate digital advertising markets. More than 100 years ago, an executive famously claimed that half of his ads were working, but lamented his inability to measure which half. The first newspaper advertising agent represented advertisers but was paid commissions by publishers (Crouse 2010). Ad blocking has affected television ad markets via the remote control, video cassette recorder and digital video recorder. Numerous media outlets have fraudulently over-reported circulation numbers to increase ad prices. Traditional advertising markets developed partial solutions; for example, the Audit Bureau of Circulation verifies print publishers' audience data. Yet all four inefficiencies manifest differently in digital advertising markets than in traditional advertising markets, and therefore are likely to require different solutions.

We write for several audiences. Many aspects of digital advertising markets remain poorly understood, so we describe opportunities for future research. We hope these might appeal to academics and also to the rapidly growing number of scientists employed by large technology companies (Athey and Luca 2019). We also write for business practitioners who may seek to understand digital advertising markets, to make individual campaigns more efficient, or to develop new ventures that can enhance market efficiency. Finally, we hope to help policy makers understand root causes of inefficiencies in digital advertising markets. Extant investigations have given less attention to the four inefficiencies discussed below than to other issues (e.g., antitrust, brand safety, disinformation, consumer privacy, market transparency, etc.). We do not take a position on whether or how governments should regulate digital advertising markets, but we hope this article will help to inform policy makers about certain key issues.

## Digital Advertising Effect Measurement

### *What is Ad Effect Measurement?*

Advertising effect measurement is the process of quantifying the incremental effect of an ad on consumer behavior. The incremental (or causal or marginal) effect of an ad is the incremental number of outcomes that were obtained as a result of the campaign, such that these outcomes would not have occurred in the absence of the campaign.

Firms measure ad effects to quantify the incremental return of their marketing investment--to determine whether their money was well spent--and to use this information to help guide future marketing decisions. Other motivations may include satisfying internal reporting requirements or benchmarking one campaign's performance against others.

Broadly speaking, campaign outcomes can be divided into two groups: branding and direct response (DR or performance). Brand advertising seeks to enhance the firm's long-term prospects by improving consumers' perceptions, attitudes, and awareness of the brand. However, quantifying these metrics, and linking them to profits, can be difficult. In contrast, a DR campaign is about driving immediate results on a conversion metric, such as purchases, store visits, registrations, downloads, or webpage visits. Our discussion of ad effects primarily involves DR campaigns, although it is important to recognize that brand advertising represents a significant portion of online ads.[2]

In many ways, advertising measurement boils down to answering a simple question: did the campaign work? And yet advertising measurement is anything but simple. The purpose of this section is to highlight common measurement challenges, to review measurement techniques, and to propose questions for future research. This review is intentionally selective and should not be viewed as exhaustive, as our strategy is to convey the challenges and methods insofar as they explain market inefficiencies and motivate suggestions for future work.

---

[2] Advertisers sometimes rely on other metrics to evaluate ad effects, such as "copy testing" metrics or social media engagement metrics. In display and search campaigns, some advertisers focus on the number of ad clicks, even though that measure is directly proportional to total campaign expenditure and may come at the expense of free clicks on organic search results.

*Measurement Challenges*

The practice of ad measurement is hardly new. However, there is a sense that digital advertising measurement presents a mix of both old and new challenges. These challenges include, but are not limited to, the following:

1. *Measurement and data availability*. Despite the volume and granularity of data available, many advertisers are still unable to connect ad exposures to outcomes at the individual level. Among other reasons, this is due to long purchase cycles, unobserved stages of consumer decision-making, and a lack of access to distribution channel members' customer data. Proving the connection between measurable outcomes, such as clicks or likes, and bottom line metrics, such as sales, is another common struggle.

2. *Strategic advertiser behavior*. Marketers often target their advertising (based on timing, characteristics, location, etc.) based on expectations of future demand, e.g., a car dealer advertising a temporary price reduction before a holiday weekend. Ad treatment often correlates with ad response--and possibly other marketing actions, such as temporary price reductions--creating a confounding correlation between ad levels and outcomes (Barajas et al. 2016).

3. *Strategic platform behavior*. Digital platforms optimize their advertising delivery. For example, if a digital advertising platform is paid per click, the platform attempts to show ads to consumers with high predicted click probabilities, creating another source of confounding variation.

4. *Strategic consumer behavior*. Consumers may pay attention to, or withhold attention from, advertisements according to ingrained habits or their valuation of the ad content (Becker and Murphy 1993; Tuchman, Nair and Gardete 2018). As a result, it may be difficult for advertisers to distinguish incremental effects of ads from consumers' baseline propensities to attend to the brand's ads. An implication is that many ad experiments recover something closer to an intent-to-treat effect rather than average treatment effects or treatment-on-the-treated effects.

5. *The complexity of ad effects*. Advertising effects may be nonlinear in the number and types of ads a consumer sees. Marginal effects of ads vary with wear-in, wear-out or weariness (Chae, Bruno and Feinberg 2019, Schmidt and Eisend 2015); competitor advertising (Danaher, Bonfrer and Dhar 2008, Shapiro 2018); and ads in other media (e.g., Naik and Raman 2003, Joo et al. 2014, Lewis and Nguyen 2015). The inability to measure all of a consumer's advertising exposures makes it difficult to obtain a fully accurate view of ad effects in many settings.

The severity of these challenges varies across campaigns, but all represent significant obstacles to quantifying ad effects reliably and accurately.

Next, we discuss the two major approaches to measuring ad effects: designing strategies to create experimental data *ex ante* or analyzing observational data *ex post*. Randomized experiments have become increasingly common as a number of online ad platforms provide such capabilities to advertisers, although usage still appears to be limited. In contrast, methods that rely on observational data are often more accessible, and thus have broad adoption by marketers. The next two subsections discuss the strengths and weaknesses of each approach with brief discussions of some representative work. The final subsection discusses some promising directions for future research.

***Experimental Studies of Advertising Effects***

Advertising experiments are widely considered the gold standard to estimate the effects of a marketing action on consumer behavior. Experiments randomly allocate units (e.g., consumers or markets) across treatment and control conditions. With sufficient sample sizes, on average the only difference between conditions is advertising in the treatment condition. Incremental conversions can be calculated by comparing outcomes between the advertising treatment condition to conversions in the no-advertising control condition.

By creating exogenous variation by design, randomized experiments address a subset of the measurement problems discussed earlier. A randomly allocated control group directly addresses the problem of strategic advertiser behavior by inducing exogenous variation in treatment. Furthermore, most online ad platforms implement experiments in a balanced manner across conditions so as to neutralize the issue of strategic platform behavior.

The academic literature increasingly uses experiments to understand advertising effects, including influential collaborations between academics and firms. The pioneering work includes a collection of papers by Randall Lewis, David Reiley, and Justin Rao, who explored the effectiveness of digital display advertising on Yahoo.com by randomizing the ad shown to different visitors (Lewis, Rao and Reiley, 2011; Lewis and Reiley, 2014).[3] This work was among the first to make clear both the benefits and difficulties of relying on large-scale online experiments for the purposes of ad effects measurement. Lewis, Rao and Reiley (2015) provides an excellent review of the early academic literature on measuring digital advertising effects.

---

[3] Field experiments in marketing go back more than 30 years, e.g. in direct mail (Bawa and Shoemaker 1987, Chapman 1986) and television advertising (e.g., Eastlack and Rao 1989, Lodish et al. 1995).

Sometimes the unit of randomization is at the geographic market level, rather than the individual consumer. In these cases, researchers have often relied on quasi-experiments or matched market tests to generate suitable treatment and control groups (Vaver and Koehler, 2011; Kalyanam et al. 2017).

A particularly influential geo-experiment was documented by Blake, Nosko and Tadelis (2015), in which eBay halted Google search advertising in a sample of cities and continued search ads in other cities. Prior to the study, eBay believed sponsored search was effective at generating incremental sales. However, the results indicated a return on advertising spending (ROAS) of negative 63%. The main reason was that consumers easily substituted to organic search results for eBay when the sponsored links were removed. This was possible because eBay's organic links ranked highly and competitors' keyword ads appeared infrequently.

eBay's market leadership position makes it natural to wonder whether these results would generalize to firms with weaker market positions. This observation motivated at least two subsequent studies. First, Coviello, Gneezy, and Goette (2017) used the same market-level research design to evaluate paid search at Edmunds.com, for which organic search results do not enjoy the same high rank as eBay. The authors found that half of paid traffic was lost when branded paid search was turned off. Second, Simonov, Nosko and Rao (2018) also found that the result did not hold similarly for most other online retailers, as competitors could poach branded keyword searchers when the focal brand does not purchase ads on its own keywords. These two papers suggest that the findings in Blake, Nosko and Tadelis (2015) might be limited to companies with a similar market position as eBay, and highlight the challenges in generalizing results from any single experiment to other settings.

Major digital platforms increasingly offer tools for advertisers to run experiments to measure incrementality on their platforms. These tools are usually offered freely, though sometimes only to sufficiently large advertisers, with the goal of helping them improve their advertising outcomes--the implicit assumption being that this will lead to increased advertising spend on the platform. A non-exhaustive list of platforms that offer experimentation tools includes AdRoll, Facebook, Google, JD.com, MediaMath, Microsoft and YouTube.[4] Although these tools are increasingly popular, widespread adoption by advertisers may take some time.

One piece of systematic public evidence about advertisers' usage of experiments comes from Simonov and Rao (2019). This paper examined online retailers' search advertising expenditures at Bing.com during the period when the results from Blake, Nosko and Tadelis (2015) were publicized in the trade press and popular media. Simonov and Rao (2019) found that 11% of firms whose branded keyword ads were not regularly purchased by competitors (similar to eBay) discontinued brand search advertising. However, the incidence of experimental advertising variation was essentially unchanged, and was uncorrelated with the size of individual firms' advertising effects. To summarize, some firms reduced their ad spend on their own branded keywords after the eBay study results were publicized; but it appears that they did so without running their own tests first.

A common problem with ad experiments is that many ad effects are small and require surprisingly large sample sizes to achieve reasonable statistical power. Lewis and Rao (2015)

---

reported the results of 25 digital display advertising field experiments with samples of 500,000 people or more. Despite large sample sizes, most experiments produced confidence intervals for ROAS wider than 100%, with the smallest confidence interval exceeding 50%. Advertising response data are highly variable, making it difficult to separate advertising's influence on conversions from unobserved factors.

The difficulty of measuring small advertising effects is a fundamental problem for advertising measurement. Whereas a product's price acts like a hammer upon consumer behavior, advertising is perhaps closer to a feather in a strong wind.

Cost is another barrier. Every consumer held out of the treatment group is a consumer who does not see the ad. If the ad is profitable, the firm may be unwilling to bear the opportunity cost of not treating consumers in the control group. Compounding this problem, some digital ad experiments serve consumers in the control group with public service announcements at the advertiser's expense. When ad effects are unknown, one might reasonably regard experiment costs as either positive or negative, depending on one's prior about the ad effects.

Lower costs of experimentation might increase the number of ad experiments that are run (e.g., Schwartz, Bradlow and Fader, 2017). Johnson, Lewis and Nubbemeyer (2017a) proposed "ghost ads" as a method to "identify ads in the control group that would have been the focal advertiser's ads had the consumer been in the treatment group." Google implemented ghost ads and has reduced experimentation costs by an order of magnitude. Johnson, Lewis and Nubbemeyer (2017b) present a meta-study of 432 field experiments at Google that used the ghost ads design. Ghost ads appears to be gaining traction within the advertising industry (Nanigans 2019).

Another approach to lowering ad experiment costs was proposed by Sahni, Narayanan and Kalyanam (2019) in the context of retargeting (i.e., targeting ads to previous website visitors). They used a single-impression public service announcement campaign to tag all consumers who were eligible for retargeting ads. This created a single control group, offering a valid baseline to estimate treatment effects of multiple varying advertising intensities.

As low-cost experimentation technologies continue to be developed and gain wider acceptance, we expect advertisers to increasingly adopt--and perhaps even demand--experimentation services on other advertising platforms.

### *Observational Studies of Advertising Effects*

Most firms either *do not* or *cannot* measure ad effects using experimental or quasi-experimental methods. Instead, most firms rely on observational data collected in the normal course of business. In such cases, academics and firms have developed many techniques to estimate ad effects from observational field data.

Probably the most common approach among firms is to analyze the market-level relationship between sales and advertising. This top-down approach is known as a Marketing Mix Model (MMM), or similarly a Media Mix Model, dating back to at least the 1960s (Borden 1964). The model normally takes the form of a time-series or panel regression of aggregate sales on aggregate marketing spending, or impressions, in each advertising medium. Additional controls include factors such as macroeconomic conditions, weather, seasonality, other marketing mix variables (e.g., price) and competitor activity. Once estimated, the model can be used to measure ROAS and to forecast sales at different levels of marketing spending in different channels.

Chan and Perry (2017) present an excellent discussion of the uses of MMMs, highlighting pitfalls and identifying opportunities for improvements. There are several arguments in favor of MMMs. First, the scope at which they operate---aggregate spending---allows MMMs to play a key role in a firm's budgeting process, helping the firm to divide its market budget across media. It would be difficult for many firms to implement a sufficient number of experiments to trace out ad effects across spending levels and media, whereas an MMM relies only on historical data. Second, the firm does not bear any opportunity costs of holding back advertising from a control group. Third, firms are more often capable of satisfying the technical and data requirements of MMMs. Fourth, MMMs can help socialize the overall concept of advertising measurement within the organization, which might help build internal support for more rigorous measurement practices in the future.

In the absence of exogenous variation, these methods rely on the assumption that they produce valid incremental effects. However, given the challenges of advertising measurement raised earlier, we should expect that some of the variation in advertising levels is not independent of the error term, as firms often divide advertising budgets across time, markets and products in expectation of future demand. Furthermore, ad spending in different media tends to be highly correlated, making it difficult to isolate channel-specific ad effects. To be clear, these concerns represent significant hurdles to the potential causal interpretation of effects obtained from observational data. But with that said, in many situations this is the most feasible path advertisers can take, so many make do with imperfect ad effect estimates to inform their decision making.

In contrast to settings with aggregate data, another common paradigm is to use individual-level data (or cookie-level data) to measure advertising effects. In these cases, advertisers often apply Multi-Touch Attribution (MTA) models, which are bottom-up approaches that seek to assign credit (to attribute) to each ad that preceded a user's conversion.[5] For example, suppose that a user clicks on a sponsored search ad, then clicks on a display ad, and then makes buys the advertised product online. To what extent did either ad contribute to generating the conversion? The most common approach is a last-click attribution model that assumes that the display ad is solely responsible for the conversion, because it occurred last. Many variations on this style of model exist and a survey indicates their usage is widespread (Forrester Consulting 2014).

A related measurement approach using individual-level data compares the conversion rates of users who were exposed to an ad campaign with users who were not exposed (e.g., Abraham 2008; often called "lift"). Lift methods rely on consumer characteristics data and model the joint likelihood of exposure and conversion to estimate the ad effect. The hope is that, after controlling for enough characteristics, the two groups are similar to the point where the ad effect can be given a causal interpretation. However, because exposure is influenced by advertiser and platform targeting strategies, conversion differences between groups may be caused by strategic behavior or unobserved characteristics, even in the absence of advertising.

Recent work by Gordon et al. (2019) demonstrates the difficulty of using observational data to estimate valid incremental ad effects. They re-analyzed 15 Facebook ad experiments comprising 1.6 billion ads served to 500 million users. Despite having rich data, they were unable to recover true ad effects accurately or consistently using observational methods. Even the sign of the bias was unpredictable: most ad effects were overestimated, but some were underestimated. As the authors put it, "even good data prove inadequate to yield reliable estimates of advertising effects."

---

[5] See Abhishek, Despotakis, and Ravi (2017) and Berman (2018) for theoretical analyses of multi-touch attribution.

Fortunately, it is possible to use observational data to address some of the advertising measurement problems. In general, quasi-experimental approaches make use of specific information about the timing of events or the data generating process to lend a causal interpretation to the estimates, and may require additional assumptions. Causality is not free.

One such approach is found in Nair et al. (2017). The data they analyzed came from a casino which had previously targeted promotions to consumers based on their observed gaming behaviors. The authors exploited knowledge of the firm's previous targeting policy to obtain unbiased estimates of consumer response to promotions. They used the estimates to segment consumers and showed, via a field experiment, that the segmentation scheme contributed to a targeting policy that generated higher incremental profits than alternate policies.

Another strategy is to take advantage of local randomization with respect to time or some other variable. Liaukonyte et al. (2015) measured changes in brand website traffic and conversions in narrow two-minute windows around the airing times of television advertisements. They based the assumption of quasi-random advertisement treatment times on a detailed understanding of television networks' sales practices, which resulted in quasi-random ordering of advertisements within commercial breaks. This helped to ensure that systematic differences between traffic and conversions in pre-ad and post-ad windows of time should be attributable solely to the presence of the TV ads. Narayanan and Kalyanam (2015) studied position effects in search advertisements by using local randomness in position when competing advertisers were very close to the threshold of winning or losing the bid for a particular position. This local randomness allows for any differences in subsequent behaviors by consumers to be attributed to the position the ad was in.

Finally, advertising exposure may be influenced by factors that are arguably independent of consumers' preferences for the brands being advertised. This is the strategy found in Hartmann and Klapper (2018), who use quasi-random geographic variation in exposure to ads during the Super Bowl to study the effects of ads in the beer and soda categories. Probability of exposure to Super Bowl ads depends on the performance of a consumer's local team. Super Bowl ads are planned at the national level many months before the event, so advertising levels should be independent of shocks to local audience sizes.

These examples demonstrate that advertisers can measure some ad campaign effects without relying on randomized experiments. But extra work is required to identify these situations, and they may not cover all the scenarios relevant to advertising decision making.

### *Future Research Opportunities*

Measuring the returns to all of a firm's digital advertising expenditures is a daunting task, especially if the goal is to do so continuously and to immediately incorporate insights into the next advertising decision. Competent, expensive, large-scale attempts to estimate ad effects have failed to measure effects with reasonable precision.

Below we outline some questions that we hope researchers consider for the future and offer some brief thoughts on how to make progress on them.

1. Advertisers can improve the usefulness of observational models. One strategy is to practice continuous experimentation across geographies, media and groups of consumers, building more random variation into the data  (Zantedeschi, Feit and Bradlow 2017). For example, some digital advertising platforms offer targeting at the zip code level, which could enable greater statistical power than traditional market-level randomizations. It would be useful to understand how firms could efficiently design this experimentation and how to incorporate it into existing models.

2. Firms can produce and leverage larger panel data sets to improve statistical power. Most advertising experiments are implemented in traditional static split/test designs. However, large panel datasets may offer increased statistical power (Berman and Feit 2018). For example, if the timing of advertising treatment can be randomized, treatment/control comparisons could potentially be made within individual consumers, thereby removing confounding individual fixed effects. A few firms can track both ad exposures and conversions in single-source customer data (e.g., Apple, AT&T, Google, Verizon, etc.). Market research firms or internet service providers are probably best positioned to create single-source panel data that measure both ad exposures and ad responses for a large number of brands, although doing so would require compliance with data privacy rules. There may also remain concerns about endogenous person/time based targeting if ads are delivered nonrandomly.

3. Advertising agencies could develop approaches that integrate ad effect measurement with advertising procurement. One such approach was described in Lewis and Wong (2018) and implemented by an agency called Nanigans; advertisement bids are manipulated to induce exogenous variation in advertisement selection. Waisman et al. (2019) addressed a similar problem at JD.com by using a multi-armed bandit approach to combine continuous experimentation with optimal exploitation of experimental findings.

4.  It may be possible to develop new quasi-experimental designs for digital advertising. A variety of such strategies have been used in other media. For example, television ad effects have recently been estimated using such instrumental variables as advertising treatment discontinuities at geographic television market borders (Shapiro 2018), exogenous shocks to TV ad demand during elections (Sinkinson and Starc 2018), exogenous TV ad insertion timing (e.g., Du, Xu and Wilbur 2019, among others), and assignments of TV networks to local cable channel numbers (Biswas, Dube and Simonov 2019). It may be possible to find and exploit similar exogenous discontinuities in digital advertising markets in order to measure causal ad effects in observational data. For example, Hill et al. (2015) show that, under particular assumptions, variation in ad viewability can help to measure ad effects, since approximately 45% of ads rendered never appear in the viewable portion of a user's screen.

5.  Firms may be able to integrate the results from observational models and randomized experiments. Some firms use experiments for tactical decisions (e.g., does this new ad design outperform the previous design?), whereas they rely on MMMs for strategic decisions (e.g., how much budget should we allocate to search versus display?). The information contained in the experiment could help inform the aggregate model, but it is not as simple as feeding the results of the experiment into the MMM. One possibility is that effects from experiments could be imposed as informative priors when estimating the MMM. Whatever the solution, many marketing measurement firms are actively searching for a unified measurement model to help deliver on this broad goal.

6. Managers may be able to embed experimentation within their profit-maximizing objective. The estimation of causal effects produces an input--the ad effect--intended to aid a manager's decision-making process. Ideally, the decision to experiment and to integrate experimental findings into profit maximization can help the manager to make use of the results for subsequent marketing decisions. Feit and Berman (2019) provide a recent example of such a framing.

7. Digital ad sellers may require techniques to validate their experimentation platforms. An increasing number of platforms offer experimentation services to help measure ad effects. Experimentation on ad platforms typically allows the advertiser to achieve greater scale, and the platform's control of the experiment should help it develop higher-quality measurement technology. However, ad platforms also face a possible credibility problem, as advertisers may worry that the platforms have an incentive to inflate ad effectiveness estimates.[6] To date, we are unaware of any systematic external audits or validations of the experimental tools that platforms provide to advertisers.

8. As more platforms offer experimentation as a service and more advertisers take advantage of this capability, it is unclear how marketers should interpret the treatment effects obtained when these effects depend on competitors' advertising policies. An experiment delivers the ad effect that is incremental conditional on all other marketing activities in the market. Lin et al. (2019) provide a framework for interpreting ad effects obtained under parallel experimentation. Future work could provide guidance on how to best generalize experimental ad effects, which may require the advertiser to form beliefs over competitors' future policies.

---

[6] There are also challenges and conflicting incentives if platforms try to estimate interactions between ads on competing platforms. For example, it would be difficult for Google to credibly offer advertisers a tool to estimate interactions between Google ads and Facebook ads. See, e.g., https://marketingland.com/where-is-google-attribution-256098

In summary, advertising effect measurement can increase advertising profits but good solutions are often unavailable. Ad effects are typically small and conversions are highly variable, so conclusive experiments are costly and rare. Observational data are plentiful but statistical analyses often do not uncover causal effects because advertisements are allocated strategically in ways that correlate with treatment. Although many people may share the intuition that it should be possible to design large-scale studies to systematically estimate ad effects, it is not yet clear when that goal will be achieved.

## Organizational Frictions and Inefficiencies

Classical economic theories of market efficiency assume that purchasers allocate budgets to maximize their own welfare. However, markets for digital ads are intermediated by specialists whose self-interested actions may distort advertising decisions.

Principal/agent problems and moral hazard are well-known sources of economic inefficiency. Generally speaking, asymmetric information enables agents to extract inefficient information rents from principals (Laffont and Martimort 2001). Perhaps the best known example is the dead-weight loss due to double-marginalization in retail pricing, which results from the divergence of manufacturer and retailer interests in a distribution channel. However, contracting problems have received limited attention in the specific context of digital advertising markets. For example, many marketers ask digital advertising agencies to evaluate their own performance; under what conditions and contracts are agencies incentivized to report truthfully?

We discuss two types of organizational inefficiencies arising in the digital advertising ecosystem. Intra-firm inefficiencies may result from misalignments between corporate officers, functional departments or business units within a firm due to different incentives or strategic objectives. Inter-firm inefficiencies may occur between marketing organizations and their external agencies, between competing agencies serving the same marketing client, between complementors within a value chain, or between colluding purchasers and sellers of advertising.

### Intra-firm Inefficiencies

Chief Financial Officers often set marketing budgets and review marketing financial performance, a common source of tension within a firm (McKinsey 2013). Many finance executives believe that marketing executives should be able to measure return on advertising spend (Fitz 2018). As explained previously, such beliefs may be misplaced. Misalignment of internal beliefs and incentives may distort marketing tactics. Marketing managers may hire outside consultants to shift blame away from themselves in the event of poor or unmeasurable outcomes. For example, they may use retargeting campaigns, which send advertisements to shoppers who previously viewed the firm's website and therefore have a higher expected baseline propensity to purchase. This creates a positive correlation between ad exposure and conversion. Lacking better causal metrics, managers may incorrectly present inflated lift metrics derived from correlations (Lambrecht and Tucker 2013; Sahni, Narayanan and Kalyanam 2019).

Another potential unintended consequence of an internal desire to measure ad effects is a distortion between short-run and long-run objectives. It is seldom possible to reliably estimate advertising effects on long-run outcomes, such as brand attitudes (Du, Joo, and Wilbur, 2019). In contrast, direct response campaigns focus on short-run actions that are easier to reliably link to ad exposure. Tension between long-run and short-run objectives is difficult to quantify, but it is possible that under-investing in brand-building advertising could reduce long-run profits. Also pertinent in this context is the potential misalignment between the firm's time horizon and that of the manager, with the manager often taking a more short-run view than the firm as a whole.

A third type of intra-firm inefficiency may result from misalignment between functional groups within the organization. Previous research has shown that poor integration of marketing and sales teams may lead to suboptimal advertising policies and customer focus (Smith, Gopalakrishna and Chatterjee 2006; Homburg, Jensen and Krohmer 2008). Marketing goals may conflict with other functional groups as well, such as when the marketing objectives are misaligned with those of the procurement department. For example, procurement may use a reverse auction to award a contract to an advertising agency in order to meet a goal of minimizing expenditures. The low-bidding agency may then provide a lower quality of service or fail to fully realize marketing objectives (Neff 2015).

Finally, distinct business units, brands or campaigns within the same firm may enter rivalrous bids on the same advertising inventory. For example, Narayanan and Kalyanam (2015) showed that after a large retailer acquired three of its rivals, the four brands continued to compete with each other by entering rivalrous bids in advertising keyword auctions.

### *Inter-firm Inefficiencies*

Many digital advertisers have replaced the traditional agency-of-record model with a complex array of external agencies. When a consumer requests a web page, any subset of the following players may be involved in delivering each digital display ad impression: publisher, ad server, supply-side platform, ad network, ad exchange, demand-side platform, multiple data management platforms, third-party verifiers, ad agency, and finally the advertiser (Choi et al. 2019). Each agent takes a cut of the advertising transaction. ANA (2017) concluded that "58 cents of each dollar ultimately purchased media inventory and audience exposure from a publisher, with 42 cents of programmatic investment consumed by supply chain data and transaction fees." Therefore, the share of digital ad spend going to intermediaries is nearly triple the traditional 15% agency commission, although it is split between many more entities in the programmatic advertising supply chain.

In addition to the programmatic marketplace, advertisers can purchase guaranteed inventory. In such transactions, the advertiser specifies the parameters of the purchase (e.g., targeted demographics, number of ads to be delivered, time frame, etc.) and the agency quotes a price. However, K2 Intelligence (2016) documented widespread obfuscation of agency mark-ups in such transactions. Arbitrage may reduce market efficiency due to moral hazard and asymmetric information between advertiser and agency, although agencies have countered that they can add value to low-cost advertising inventory by applying analytics.

Another type of inefficiency may arise when advertisers fail to coordinate agencies whose work generates spillovers across marketing channels. For example, a large body of research has found significant spillovers between advertising placements in traditional media and digital activities such as brand search, website traffic, online sales and social media conversations.[7] Few marketers are known to coordinate their advertising activity to account for such spillovers. An integrated channel would consider the effect of traditional advertising on volume of online search, types of keyword searched, search advertising expenditure and sales, and would quantify spillovers to allocate advertising budgets across media (Kim and Balachander 2017). However, the typical approach is for marketers to hire specialist agencies for each individual medium (e.g., traditional advertising, search engine marketing, social networking, search engine optimization, website analytics, etc.), each of whom operates independently and competes for a larger share of the marketer's advertising budget.

---

[7] See, e.g., Dinner et al., 2014, Fossen and Schweidel 2017, Joo et al. 2014, Lewis and Reiley 2013, Liaukonyte et al. 2015, Du, Xu and Wilbur 2019, among others, for effects of TV ads on digital behaviors. Lewis and Nguyen (2015) document spillovers from digital display advertising to search behaviors.

Another type of inefficiency related to advertising agencies is how to properly align incentives to produce efficient advertising creative content. On one hand, advertising agencies' creative ideas can be stolen during the pitch process, especially when clients work with multiple agencies (Dan 2014). Therefore, some agencies may not reveal the full idea during the contracting process, potentially leading to adverse selection or under-investment in idea production (Horsky, Horsky and Zeithammer 2016). On the other hand, advertisers have difficulty evaluating the quality of agencies' creative strategies due to the complexity and high dimensionality of the messaging problem. Although messaging content is still largely driven by human intuition, there is increasing movement toward algorithmic production and validation of advertising messages in digital environments.

Three recent trends affect the market for advertising agency services. First, there is a long-term trend toward in-housing, in which marketers create internal advertising agencies. In-housing is used primarily when firms have internal creative abilities or straightforward advertising requirements (Horsky 2006). ANA (2018) surveyed large advertisers, finding that 78% of advertisers had in-house agencies in 2018, up from 42% in 2008; and also that 90% of those advertisers continue to work with external agencies in addition to their internal teams. Second, digital platforms have simplified their customer interfaces, a move which may encourage some advertisers to forego agency services. For example, Google recently made "smart display campaigns" the default interface to Google Ads. The streamlined purchase process relieves the advertiser of control over bidding, ad placement, user targeting and final control over the ad creative. However, advertisers using this interface may pay advertising costs that exceed advertising revenues, as they no longer can enter a maximum bid per click. Third, there is an emerging trend toward large advertisers requesting log-level files (i.e., impression-level data) from supply chain partners such as advertising exchanges and publishers (Sluis 2019). Data sharing enables advertisers to monitor partner agencies and suppliers, trim wasteful spending and find undervalued inventory.

A different inter-firm inefficiency may occur when advertisers compete in advertising auctions with complementors within the same value chain. For example, Hilton and Expedia both sell Hilton rooms in New York, and Expedia takes a commission when it sells a Hilton room, yet both place search ads on keywords like "hotel New York." Cooperative advertising policies may even exacerbate this competition; Hilton may subsidize the advertising expenditures of Expedia, even though this directly increases Hilton's keyword advertising costs by making the keyword auctions more competitive. Cao and Ke (2019) and Jerath et al. (2019) examine such practices.

Finally, there are inefficiencies which may increase advertisers' profits at the expense of overall market efficiency. One such effect may occur when competing advertisers use a common advertising agency, as doing so may coordinate marketing policies or facilitate information sharing (Villas-Boas 1994). Another may occur when the structure of programmatic advertising marketplaces facilitates collusion between purchasers. Digital advertisements are sold in high-frequency auctions with repeated contact among bidders, creating ripe conditions for collusive bidding strategies and punishments for defection. Decarolis and Rovigatti (2019) found that advertising agency consolidation is associated with decreased costs in keyword auctions when merging agencies represent competing bidders. There has been a recent trend toward global consolidation of advertising agencies (Pathak 2018), suggesting that such effects may be considerable.

### *Future research opportunities*

There are a number of important research opportunities within this area.

1. Recent theoretical work in information economics has introduced Bayesian persuasion models and the analysis of strategic information design (Kamenica and Gentzkow 2011, Bergemann and Morris 2016). Bayesian persuasion models characterize optimal informational strategies (i.e., what message to send, to whom, and when or how much information to reveal) to influence receivers' decisions. This seems to be a promising theoretical framework on which to build an understanding of optimal advertising strategies for advertising content, targeting, and platform design (Mayzlin and Shin 2011; Fuchs et al. 2016; Zhong 2018).

2. Principal/agent and theory of the firm-style analyses have rarely been applied to advertising agencies. It may be that such settings can be understood with straightforward applications of canonical models. However, given the particular institutional details, we suspect there is substantial scope to extend existing theories in the process of applying them to advertising phenomena. Future research could include normative guidelines about how optimal contract terms depend on features of a market and the contracting parties. Contract terms likely affect incentive alignment, effort elicitation, payment models, in-housing/outsourcing decisions, and agency selection and coordination.

3. There are opportunities to further develop and improve techniques to optimize digital advertising content and context. Machine learning has demonstrated progress in identifying sentiments (Cambria et al. 2017) across a wide variety of unstructured data including text (e.g., Netzer et al. 2012), images (e.g., Liu et al. 2018) and video (e.g., Pérez-Rosas et al. 2013). The meaning of advertising content depends on the context (Poria et al. 2017, Rafieian and Yoganarasimhan 2019), such that algorithms are needed to detect and appropriately understand the context and work with multimodal data (Soleymani et al. 2017). Similar opportunities may be available when a marketer can access log files from multiple partners, as the data may be used to optimize advertising frequency, quantify the impact of context on conversion, optimize procurement techniques or evaluate ad agency services.

4. Perhaps the most promising area for research involves designing auction and advertising allocation systems that more effectively align the incentives of the parties involved (e.g., Hu, Shin and Tang 2016). Johnson and Lewis (2015) propose a cost-per-incremental-action (CPIA) pricing model for advertising, with the goal of aligning all participants' objectives on profit maximization for the advertiser, thereby reducing the misaligned incentives that exist in common advertising pricing models.

Additional research is needed to test such models in the field and to adapt them to various digital advertising markets.

## Ad Blocking

### *Description and Background*

Early digital advertising efforts developed intrusive formats such as pop-up ads or autoplaying audio/video ads. This led to consumer demand for ad blockers, applications that allow users to passively block advertising from showing up in their browsers. Most ad blockers came in the form of free-to-use browser extensions enforcing a set of community-defined rules for ads. Recent estimates of ad blocking prevalence vary from 8% to 47% (Searls 2019). In mid-2019 Google incorporated ad blocking features into its market-leading Chrome browser, while continuing to enable third-party ad blockers.

The phenomenon of ad blocking highlights an important tension in the current digital advertising ecosystem. Consumers are flooded by ads, most of which are not interesting or relevant. Prior to ad blocking technology, consumers had no choice but to tolerate the ads; most users were not willing to pay for most content, so content providers relied on advertising revenues. Ad blocking threatens this model, as advertisers' most sought after consumer segments are often the most likely to install ad blockers.

The most popular ad blockers work by intercepting browser requests to lists of known ad servers, so advertisers are not charged for blocked ads. However, some ad blockers are more aggressive: they only block ads after they are requested, thereby wasting the advertiser's money.

One ad blocker, AdNauseam, even *clicks* on all blocked ads, in order to inject noise into the user data that underpins digital advertising.

Platforms such as web browsers and mobile operating systems enable or restrict ad blocking services. However, platforms that sell ads have conflicting incentives. Their revenues depend on consumer experience but also advertising exposures. It is thus not surprising that critics argue that Google blocks relatively few ads and interferes with competing ad blockers. Mobile operating systems assert more control over applications than traditional desktop operating systems. Apple and Google have rules about apps that interfere with how other apps work, making it hard to offer blocking of in-app ads.

Ad-supported publishers are perhaps most harmed by ad blocking. Facebook and some other sites deliver ads and content from the same servers, preventing ad blockers from identifying which page elements are ads. Other publishers seek to detect ad blocker usage and withhold content from consumers who do not allow ads; Zhu et al. (2018) found that 30% of the top 10,000 websites detect ad blockers. Some publishers, including Facebook, invest heavily in website engineering to circumvent ad blockers and unblock ads. Other publishers explicitly request that users whitelist the site, i.e. selectively disable their ad blocker, so the site's ads may be displayed. Software developers have responded with ad blocker-blocker-blockers, which enables the consumer's browser to obfuscate ad blocker usage, thereby preventing publishers from withholding content.

The overall effect of ad blocking on advertisers, consumers and publishers is unclear. Using the circulation spiral theory of newspapers (e.g., Gabzewicz, Garella and Sonnac 2007), one might predict that ad blockers will put advertising-supported media out of business. By reducing ad revenues, ad blockers increase direct-access prices and thereby reduce publishers' incentives to produce high-quality content. Taken to its extreme, ad blockers would destroy the ad-supported internet and thereby become irrelevant, as there would be no more ads to block. A contrary argument holds that ad blocking serves consumer interests: some ad blockers even positioned themselves as a consumer movement (Katona and Sarvary 2018). Ad blockers say they only block the most intrusive ads and that users can whitelist any site they want. In fact, ad blockers whitelist some sites by default.

This brings us to an important question: how do the ad blockers make money? Most ad blockers do not directly charge users for downloads or ad blocking services. The typical model used by ad blockers is to demand payment from publishers in exchange for whitelisting-by-default. If the publisher pays and conforms to some guidelines about ad formats and intrusiveness, the ads will be displayed to ad block users by default. Large publishers typically pay 30% of ad revenue that would otherwise be blocked; small publishers can be whitelisted for free if they agree to follow programs such as the "acceptable ads" standard.

Publishers have argued that such practices amount to extortion. There have been numerous lawsuits. Some of the most important cases have taken place in Germany where the company that makes Adblock Plus, Eyeo, is based. The German Supreme Court has ruled that ad blocking and the practice of soliciting payment for whitelisting is legal (Ha 2018).

### *Future Research Opportunities*

The phenomenon of ad blocking raises a range of future research opportunities.

1. It is interesting to study how consumers trade off the intrinsic benefits of content consumption with advertising quantity, advertising nuisance and subscription prices.

Huang, Reiley and Riabov (2018) ran experiments that showed that increased advertising on Pandora.com directly decreased consumers' site usage and increased consumer subscriptions. It would be interesting to measure similar effects in other contexts; to condition on the nuisance level of the ads served; and to estimate effects of ads on adoption dates of ad blockers, among other relevant outcome variables. It also would be interesting to study how design elements within publisher requests for whitelisting might correlate with response metrics.

2. There may be better ways for browsers and ad blockers to improve mechanisms to allocate attention. On one hand, browsers that specialize in privacy and ad blocking have not realized high market shares. On the other hand, there are some notable experiments ongoing. For example, the browser Brave implements the "Basic Attention Token" to reward users for looking at ads and send money to publishers the user likes. New designs might take hints available from recent literature on privacy, which shows that consumer valuations of privacy respond strongly to framing and context (Acquisti, John and Loewenstein 2013, Winegar and Sunstein 2019).

3. Recent work by Shiller, Waldfogel and Ryan (2018) finds that, as a larger proportion of a site's visitors use ad blockers, the site's quality declines. Further research is needed. How does a site's revenue loss from ad blocking compare with payments to ad blockers? When do sites switch from ads to paid subscription models, unblockable "native advertising" or other blocking-proof business models? Similar questions can be asked for advertisers, especially those whose target customer segments are most likely to use ad blockers. How do ad blocking and whitelisting affect ad placements and ad prices?

4. It is interesting to understand how ad blocking changes product market outcomes. There are a few recent theoretical pieces on the interaction between publishers and ad blockers (Gritckevich et al. 2019) and differentiation between publishers (Despotakis et al. 2017), but there are more directions to explore. For example, how do ad blockers compete with each other? Blocking more ads than a rival ad blocker may attract more consumers, and it may also impact the value to publishers of paying to be whitelisted by default. It would be interesting to quantify these tradeoffs, as they likely rely on the extent of consumer multihoming, both across publishers and across ad blockers.

5. There is also the question of how advertisers react to ad blocking. Chen and Liu (2019) argue that ad blocking can incentivize higher quality ads when ads send informative signals. Other fundamental questions should also be addressed. If ad blocking reduces the overall supply of advertising space, that must raise the overall price of reaching customers. The increase could be disproportionate for more ad-averse consumers. How does that impact advertisers? Intuitively, the direct effect should be negative, but perhaps limited advertising also reduces price competition (Dukes and Gal-Or 2005) or increases the effects of ads that are not blocked. Future research could dig deeper to examine how other dimensions of advertising content are affected by ad blocking. If a brand has to pay more to reach a consumer, the brand might change how much it focuses on communicating a single piece of information (e.g. price, product feature, etc.) or alter the mix of entertaining and informative content contained in the ad.

6. Platforms' business models may incentivize them to set ad blocking defaults or limits in mobile operating systems and desktop browser software. Overall, it seems likely that platforms selling ads will restrict ad blocking more than platforms that charge consumers directly for devices and software, in order to limit harm to advertising revenues.

## Digital Advertising Fraud

### Description and Background

Digital advertising fraud is a collection of practices that misrepresent advertising inventory or disguise machines as humans in order to steal advertising budgets. It is fundamentally difficult or impossible to measure, but it appears to be widespread.[8] eMarketer (2019b) reported that recent estimates of fraud vary from $6.5 to $19 billion, and that fraudsters target high-price ad impressions, new markets and new media. Adobe reported in 2018 that about 28% of website traffic showed "strong non-human signals." The IAB Tech Lab reported that "just 59.8% of clicks could be confirmed as human traffic" (AdWeek 2019). Pixalate (2018) estimated that 10-15% of programmatic desktop advertising was invalid traffic. Davies (2019) reported that "28% of global mobile media budgets are wasted on fraud." Sweeney (2019) reported that 95% of marketing executives surveyed said that digital media must become more reliable and 21% said they have cut ad spend due to inaccurate, questionable or false reporting.

We know of six basic motivations to commit advertising fraud.

1) Publishers may over-report or misrepresent audience metrics to increase ad revenues.

---

[8] Skepticism is appropriate when interpreting fraud measurements. Fraud is defined by intention, which is not directly observable in ad data. Indirect measures of ad fraud trade off subjective risks of false positive observations of fraud against false negative detections of valid ad exposures. Fraud detection firms might over-report fraud to attract business; ad sellers might under-report fraud to reassure clients.

2) Advertisers may click competitors' ads to harm rivals' market inferences, ad budget or brand awareness; or to reduce competition in ad auctions by depleting a rival's budget.

3) A firm may commit detectable advertising fraud that ostensibly benefits a rival, with the goal of inducing a platform or other intermediary to punish the rival.

4) Advertising market intermediaries, including agencies, ad networks, demand-side platforms and supply-side platforms, may misrepresent the inventory or the bids they have available in an effort to alter ad prices, ad sales, or commissions. They also may deviate from contracts to manipulate ad auctions, as in the recent bid caching scandal.

5) Publishers or agencies may distribute advertisements surreptitiously or insert false information into advertisers' conversion tracking systems to claim undue credit from advertisers who pay publishers per conversion achieved.

6) Firms or individuals may create fraudulent profiles on social networks to falsify measures of influence, ad clicks, or seemingly organic discussions (astroturfing).

Collectively, there are dozens of known schemes to commit ad fraud.

Fraud techniques that use machines to mimic human behavior are often enabled by "botnets," which consist of a central server and a host of malware-infected computers. The server directs connected machines to take specific actions in ways that resemble their owners' organic behaviors. Machine activity is therefore difficult to distinguish from human activity.

Security researchers have publicized several *ad hoc* botnet discoveries. For example, the discovery of the 3ve botnet in 2018 led to the first criminal charges filed for advertising fraud.

The indictment stated that "more than 1.7 million infected computers… download[ed] fabricated webpages and load[ed] ads..."[9]

Another set of fraud techniques misrepresents advertising inventory, as several publishers have shown. The *Financial Times* did not sell inventory in programmatic advertising marketplaces, but it found fraudulent FT.com ad inventory offered in 25 ad exchanges, with fraudulent video inventory exceeding truly available inventory by a factor of 30. *The Guardian* purchased intermediaries' Guardian.com video inventory on open advertising exchanges, reporting that 72% of ads purchased falsely claimed to occur on Guardian.com.

Large digital platforms have been inconsistent in their claims about digital advertising fraud. Google states in its advertiser-facing webpages that "we protect you from invalid activity and advertising fraud" and describes a variety of approaches. However, internal communications filed as evidence in a 2017 lawsuit indicated that Google recalled fraudulent funds from publishers but lacked an internal system to return those funds to defrauded advertisers, and that Google's advertiser contracts did not require the company to offer refunds to defrauded advertisers (AdTrader 2018). Facebook states that "we believe fake accounts are measured correctly within the limitations to our measurement systems." However, Facebook does not specifically define fake accounts, although its community standards do prohibit "accounts that are fake" without elaboration. Facebook's Community Standards Enforcement Reports show that

[9] The indictment further stated "To create the illusion that real human internet users were viewing the advertisements loaded onto these fabricated websites, the defendants programmed the datacenter servers to simulate the internet activity of human internet users: browsing the internet through a fake browser, using a fake mouse to move around and scroll down a webpage, starting and stopping a video player midway, and falsely appearing to be signed into Facebook. Furthermore, the defendants …[made] it appear that the datacenter servers were residential computers belonging to individual human internet users who were subscribed to various residential internet service providers." https://www.justice.gov/usao-edny/pr/two-international-cybercriminal-rings-dismantled -and-eight-defendants-indicted-causing

Facebook "took action on" 6.6 billion fake accounts in the 12 months ending in September 2019, but did not specify the actions taken. Facebook has restated advertising metrics numerous times (Peterson 2017).

There are two types of efforts to address digital advertising fraud. One type of effort increases market transparency and accountability to reduce the incidence of advertising supply chain participants from stealing from each other. For example, the IAB Tech Lab defined the ads.txt standard in 2017 to enable publishers to publicly identify authorized advertising sellers and resellers, so that buyers can audit ads.txt files to avoid unauthorized sellers. Publisher adoption is widespread. Newer standards ads.cert, sellers.json and the OpenRTB Supply Chain Object enable similar disclosures by other market participants. Anecdotally, it appears that these efforts have meaningfully reduced fraud borne by supply chain participants.

The other type of anti-fraud effort helps advertisers to avoid paying for fraudulent ads *ex ante* and to seek reimbursement for fraud detected *ex post*. Most importantly, there are a range of firms that specialize in ad fraud detection, including some that partner with large digital advertising platforms to directly analyze some data. There are also industry collaborations, such as the Trustworthy Accountability Group, which publishes a monthly Data Center IP List, a "common list of [Internet Protocol] addresses with invalid traffic coming from data centers where human traffic is not expected to originate." There are also industry-standard contractual language intended to help advertisers obtain relief for fraudulent ads. Measurement difficulties make it challenging to know how advertiser-facing fraud has changed over time. Advertiser concern and awareness about fraud have increased, but at the same time, fraudsters have grown more sophisticated and discovery of large-scale botnets has increased.

The issues underlying digital ad fraud are nontransparency and nonverifiability of human recipients of ads. Advertising delivery systems seldom require proof of humanity; for example, publishers do not require users to solve a reCAPTCHA before serving ads. Ads are delivered when a computer requests a page from a web server, placing ad exposures in the set of activities that are trivial for botnets to perform programmatically. As Vinton Cerf, co-creator of the internet protocol, put it, "We didn't focus on how you could wreck this system intentionally." Digital ad fraud is often described as a cat-and-mouse game in which fraudsters develop new tactics when previous tactics have been neutralized.

We emphasize the role of botnets because evasion of detection is of central importance to those who would commit advertising fraud. Fraud that can be detected can be reversed, with charges reverting back to the client marketers and fraudsters being excluded from established advertising networks. Therefore, competent ad fraud is, by definition, fraud that is sufficiently unlikely to be detected. Major advertising networks say that they seek to detect advertising fraud, but they do not fully explain the specific techniques they use. Fraud detectors are hamstrung in gaining advertisers' trust: sharing their precise fraud detection algorithms could help unscrupulous actors to avoid fraud detection. There is also a credibility problem: Wilbur and Zhu (2009) proved that, under general conditions, digital advertising platforms' revenues may rise when they fail to detect ad fraud, even when advertisers adjust their bids based on expected fraud levels. To summarize, small advertisers are forced to trust in partners' efforts to detect fraud, even though the partners get paid every time their detection algorithms fail.

***Future Research Opportunities***

The academic literature on ad fraud is relatively sparse. There have been a few theoretical studies examining how advertising market conditions and business models impact various players' incentives to engage in fraud (e.g., Chen et al. 2015). There is a more sizable literature in computer science proposing algorithms to detect particular fraud schemes (e.g., Behdad et al. 2012). Finally, there are empirical papers that relate fraud prevalence to other market conditions in order to offer guidance to firms. For example, Edelman and Brandi (2015) relate marketers' monitoring efforts to online affiliate fraud. They found that monitoring by outside specialists is more effective at punishing clear violations of marketer rules, whereas internal marketing staff are more effective at punishing borderline rule violations.

More research is needed. We see five areas as particularly important.

1. Theoretical analyses have seldom considered the economic antecedents and consequences of digital ad fraud. We think it would be interesting to model digital advertising as a credence good as many advertisers cannot reliably measure ad effects. An open question is how credence goods markets function when delivery of the service is only partially verifiable, and how optimal contracting terms depend on the properties of the transaction. A related question is whether there are strategies buyers or market makers can adopt independently to improve market efficiency in such settings. Model-based predictions might help to identify opportunities to improve market regulations.

2. Another area of opportunity is to help platforms and fraud detection firms design better approaches to detect and reverse ad fraud. In general, the topic of how to attack machine learning algorithms and how to resist attack is a topic of active research (e.g., Wang et al. 2019). Consideration of adversarial methods may help firms to anticipate and defend against future developments in fraud technology. It also may help to identify opportunities to incentivize humans to prove their humanity in order to better measure when ads are being delivered to people as opposed to bots.

3. It may be interesting to adapt market designs and regulations from other contexts. For example, securities fraud may have been affected by government regulations or dispersion of trading across competing stock markets. There may also be lessons available from other examples of fraud, such as insurance fraud or identity theft. Successful efforts to detect, prevent, reverse or deter fraudulent activity in other settings may have implications for effective management of digital advertising fraud.

4. It is important to understand empirical predictors and impacts of fraud. If institutional ethics guidelines allow, researchers could simply run experiments by purchasing various types of botnet actions, as they are available cheaply online; then observe how fraudulent traffic changes market outcomes. Ideally, an approach might be developed that enables individual agencies or advertisers to audit platforms', publishers' and fraud detection firms' claims about audiences delivered and fraud prevented.

5. Another approach could use certain events as discontinuities in quasi-experimental designs, such as privacy regulation changes, publishers' adoption dates of ads.txt or other transparency standards, or major public disclosures of botnets or ad fraud schemes. Examining how market data and indicators of fraudulent activity change around event times could show how fraud responds to incentives and distorts markets.

We close by highlighting two hypotheses related to ad fraud and allocative efficiency. First, there are trade-offs between consumer privacy, market transparency and fraud detection. In particular, the more signals there are about human identity and activity, the more information is available to potentially distinguish humans from machines. Therefore, it seems possible that increasing consumer privacy may also hinder detection of ad fraud and market transparency.

Second, we think the application of blockchain technology is promising in digital advertising markets. However, widespread adoption of new standards requires coordinated action. Will the walled gardens adopt blockchain or other anti-fraud technologies? As an example, Ads.txt adoption was boosted significantly when Google adopted and promoted the standard, but Google did not actively support the standard until after a significant number of publishers adopted it. To date, several blockchain-based solutions have been proposed for digital ad markets, but we are not aware of any that have achieved significant traction. New standards may need to anticipate the chicken-and-egg problem to get ad purchasers, sellers and intermediaries on board.

## *Broader Discussion*

### *Interactions between Inefficiencies*

We have discussed each source of inefficiency in isolation, but they may interact to exacerbate market-level inefficiencies. A prominent example is when advertising agencies' private incentives depart from marketing principals' profit incentives to measure credible ad effects. This interaction could help to explain the relative scarcity of experimental evidence about advertising effects, and the frequent misinterpretation of observational methods as yielding causal ad effects.

Other harmful interactions between inefficiencies appear likely. Agencies may misreport the extent of ad blocking or ad fraud to their clients, especially if negative information might reduce the client's advertising budget and, consequently, the agency's commission. Another prominent example is that ad fraud or ad blocking may contribute to the statistical challenge of measuring ad effects by reducing the number of valid advertising exposures delivered to consumers in treatment and control cells of an experiment.

### Policy-Relevant Research Opportunities

Several high-profile regulatory reports have recommended new regulations for digital advertising markets. Those reports have described many important topics extensively (e.g., antitrust, brand safety, disinformation, consumer privacy, market transparency, etc.). However, they have covered the four themes treated in this article in less detail. We now consider the question of how research could help to inform broader policy considerations, in addition to the four individual themes treated previously.

Ultimately, digital advertising markets exist to create value for two types of agents they aim to connect: consumers and marketers. Consumers supply attention and realize utility from advertisements, media content and advertised products. Marketers buy ads and realize profits from product sales. All other relevant parties are intermediaries enabling interactions between consumers and marketers. Therefore, consumer welfare and producer profits should be the two primary concerns in designing government policy.

*Consumer-level questions*. It is particularly important to understand how advertisements affect consumer search for information about products and prices. Some advertising may enhance product-market outcomes for consumers whereas other ads may suppress product search and reduce competition in the product market. It is also important to uncover when advertising may have direct negative impacts on consumer utility, as is suggested by the large literature on advertising avoidance, and the extensive use of ad blockers. There is existing literature on these topics, but more comprehensive and systematic evidence is needed to guide policy due to the heterogeneity among consumers, advertisers, and the variety of advertising channels and messages.

*Marketer-level questions*. The ad effect measurement section explains some significant obstacles to profit-maximizing advertising decisions. An important topic that is not well documented is how different types of firms make different types of advertising decisions. Some advertisers are known to be more sophisticated and make more data-driven decisions than others, but we are aware of little public evidence about the financial returns to data-driven advertising decision-making. There are similar questions about estimating the returns to using specific new tools in the advertising process, such as automated ad content generation, programmatic advertising purchases, fraud detection and brand safety monitoring. Another open question is how different types of firms should structure their external advertising agencies and what contracting mechanisms might be optimal in what situations. Differences in advertiser objectives, capabilities, ownership structures, and firm size may help to explain why some firms make more data-driven decisions than other firms. Overall, there is substantial scope to better understand the antecedents and consequences of advertiser behavior.

*Industry-level questions*. There are not many examples of successful regulation in multi-sided platform industries like digital advertising markets. It is unclear how to optimally trade off conflicting goals among various players. One relevant trade-off may occur between consumers' interests, such as privacy, and advertisers' interests, such as fraud detection. Another relevant trade-off may occur between large platforms' interests in acquiring early-stage companies and potential competitors' incentives to enter and innovate.

It would also be interesting to consider what metrics regulators can use to measure the efficiency of digital advertising markets. It seems likely that regulations that entrench incumbents may deter potential entrants, so entry and innovations might be potential indicators of well-functioning markets. It may be that ad blocking installations decrease with market performance for consumers and therefore might underpin relevant metrics. Perhaps advertiser usage of experimentation platforms, and subsequent changes in advertising spending, could help to indicate how digital advertising markets are serving advertisers' interests.

Finally, there is a question of whether regulatory regimes will be robust to platforms' active preparations to be regulated. For example, some technology companies have been anticipating antitrust regulation for years (Athey and Luca 2019). If regulatory policy is not robust to adversarial interference, it could exacerbate the problems it seeks to resolve.

## Conclusion

Digital advertising markets offer numerous innovations over traditional advertising. Perhaps most notable is the immediacy with which a small business can target and advertise to local consumers, along with the increased variety of advertising formats, relevance, market structures and interactivity. However, as with most technological innovations, the shift to digital advertising has produced costs as well as benefits. We have reviewed four common issues that are likely to hinder allocative efficiency in digital advertising markets. Most marketers either do not or cannot measure incremental ad effects; uncertainty about ad effects may distort market demand for advertising. Numerous intermediaries separate marketers from publishers, each of whom takes a cut of advertising expenditures and has its own private information and incentives, leading to asymmetric information and moral hazard. Consumers use ad blocking software to passively prevent advertisements from being displayed, resulting in misappropriated advertising revenues and reduced incentives to provide media content. Advertising fraud misrepresents advertising opportunities and directs ad exposures to machines to steal advertising budgets.

We hope the discussion is helpful to academics, scientists employed by advertising companies, policy makers and executives as they seek to understand modern digital advertising markets. We do not claim that these four issues are necessarily more important than others; we have focused on them because we believe that they are less well understood than other topics that have received more careful study. Such topics as antitrust, brand safety, disinformation, consumer privacy, and market transparency are also important policy considerations. We have not taken a position on whether or how digital advertising markets should be regulated, but we do prefer policy makers to have full information and hope this article may be helpful toward that goal. Overall, we hope this survey of market inefficiencies in digital advertising markets might help to develop scientific literature and better inform the policy making process. Many opportunities remain and much work has yet to be done.